\documentclass[amsmath,floatfix,twocolumn,superscriptaddress]{revtex4-1}
\usepackage{amssymb,amsmath}
\usepackage{graphicx,array}
\usepackage{dcolumn}
\usepackage{psfrag}
\usepackage{bm}
\usepackage{color}
\usepackage{epstopdf}

\begin{document}
\title{Colossal room-temperature electrocaloric strength aided by hydrostatic pressure 
       in lead-free multiferroic solid solutions}

\author{C\'{e}sar Men\'{e}ndez}
\affiliation{School of Chemistry, The University of Sydney, NSW 2006, Australia}

\author{Claudio Cazorla}
\thanks{claudio.cazorla@upc.edu}
\affiliation{Departament de F\'isica, Universitat Polit\`ecnica de Catalunya, 08034 Barcelona, Spain}

\begin{abstract}
Solid-state cooling applications based on the electrocaloric (EC) effect are particularly promising 
from a technological point of view due to their downsize scalability and natural implementation in 
circuitry. However, EC effects typically occur far from room temperature, involve materials that contain
toxic substances and require relatively large electric fields ($\sim 100$--$1000$~kV~cm$^{-1}$) that 
cause fateful leakage current and dielectric loss problems. Here, we propose a possible solution to 
these practical issues that consists in concertedly applying hydrostatic pressure and electric fields 
on lead-free multiferroic materials. We theoretically demonstrate this strategy by performing first-principles 
simulations on supertetragonal BiFe$_{1-x}$Co$_{x}$O$_{3}$ solid solutions (BFCO). It is shown that 
hydrostatic pressure, besides adjusting the occurrence of EC effects to near room temperature, can 
reduce enormously the intensity of the driving electric fields. For pressurized BFCO, we estimate 
a colossal room-temperature EC strength, defined like the ratio of the adiabatic EC temperature change 
by the applied electric field, of $\sim 1$~K~cm~kV$^{-1}$, a value that is several orders of magnitude 
larger than those routinely measured in uncompressed ferroelectrics.     
\end{abstract}
\maketitle

One of the limiting factors of modern microelectronic devices is their tremendous heat dissipation
density, which needs to be mitigated through cooling in order to ensure proper performance. Current 
refrigeration technologies, however, rely on compression cycles of environmentally harmful gases and 
cannot be scaled down to microchip dimensions. Electrocaloric (EC) cooling is a highly promising solid-state 
refrigeration technology for thermal management of chips and microcircuitry owing to its high efficiency,
environmental friendliness, and easy miniaturization \cite{liu16}. EC refrigeration exploits the reversible 
thermal change of ferroelectric materials resulting from phase transitions induced by external electric 
field variations. Large EC isothermal entropy changes, $\Delta S_{\rm EC}$, of $\sim 10$~J~K$^{-1}$kg$^{-1}$ 
and adiabatic temperature changes, $\Delta T_{\rm EC}$, of $\sim 1-10$~K have been measured in ferroelectric 
materials like BaTiO$_{3}$ \cite{bto1,bto2}, Pb(Zr,Ti)O$_{3}$ \cite{ec1} and HfO$_{2}$ \cite{ec0}, to cite 
few examples. 

Nonetheless, unfortunately, the largest EC effects observed to date normally occur at temperatures far 
from room temperature \cite{problem1}, involve materials that contain toxic substances like lead and 
require large electric fields that are energetically costly and produce adverse leakage currents and 
dielectric losses \cite{problem2,problem3}. Recently, several materials design strategies have been proposed 
to overcome these common EC problems. For instance, by exploiting electrostatic coupling and interface effects 
in lead-free ferroelectric relaxor heterostructures, an unprecedentedly large EC adiabatic temperature shift 
of $\approx 23$~K has been realized near room temperature for moderate electric bias ($\varepsilon_{c} \sim 
100$~kV~cm$^{-1}$) \cite{ec6}. Nevertheless, the magnitude of such EC effects can be strongly influenced by 
the specific details of the heterostructure synthesis process and thus in practice $\Delta T_{\rm EC}$ may strongly 
fluctuate from one sample to another. Another recent EC advancement has been reported for the layered hybrid perovskite 
ferroelectric [(CH$_{3}$)$_{2}$CHCH$_{2}$NH$_{3}$]$_{2}$PbCl$_{4}$ \cite{ec7}, in which a sharp first-order 
ferroelectric phase transition associated to a high-entropy change occurs instead of the continuous phase 
transformation associated to a low-entropy change that is characteristic of inorganic ferroelectric perovskites \cite{cohen92}. 
In this case, a giant $\Delta T_{\rm EC}$ of $11.1$~K has been measured at room temperature for a small 
electric field of $29.7$~kV~cm$^{-1}$. However, the implicated material still contains lead and the degree 
of reversibility associated to such giant EC effects appears to be quite limited. 

In this work, we propose a completely different approach for the enhancement of EC effects that consists
in the application of multiple external fields on lead-free multiferroic materials able to undergo sharp 
first-order phase transitions. In particular, we demonstrate by means of computational first-principles 
methods that the sequential operation of hydrostatic pressure and electric fields in BiFe$_{1-x}$Co$_{x}$O$_{3}$ 
solid solutions (BFCO) can trigger large and inverse EC effects of $\Delta S_{\rm EC} \approx 5$~J~K$^{-1}$kg$^{-1}$ 
and $\Delta T_{\rm EC} \approx -5$~K at room temperature. Moreover, aided by pressure BFCO displays a 
colossal EC strength of $\sim 1$~K~cm~kV$^{-1}$, defined like $|\Delta T_{\rm EC}|$/$\varepsilon_{c}$, 
which surpasses by several orders of magnitude the typical values reported for uncompressed ferroelectrics.

\begin{figure*}[t]
\centerline
        {\includegraphics[width=1.00\linewidth]{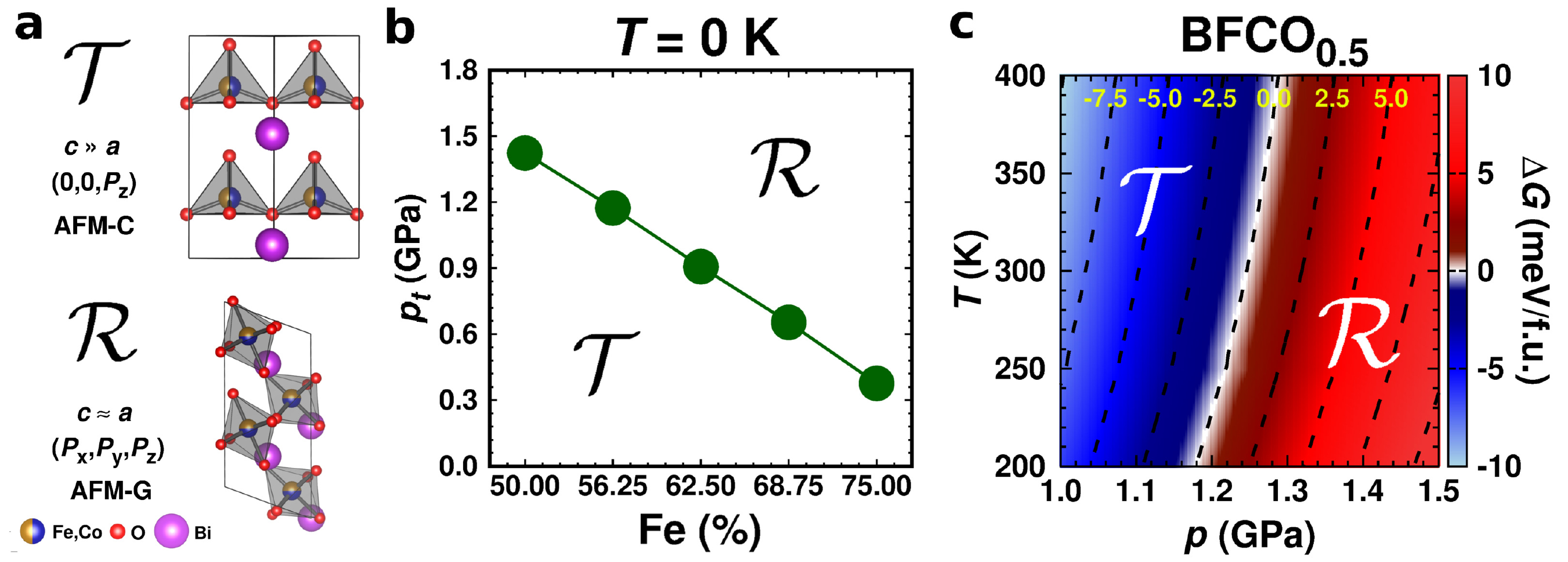}}
	\caption{{\bf Phase competition in BFCO under pressure.} {\bf a}~Sketch of the competitive tetragonal ($\cal{T}$)
	and rhombohedral ($\cal{R}$) multiferroic phases. The corresponding electric polarization and antiferromagnetic 
	spin ordering are indicated. {\bf b}~The $\cal{T} \to \cal{R}$ transition pressure calculated at $T = 0$~K and 
	disregarding likely quantum nuclear effects expressed as a function of composition. {\bf c}~First-principles 
	$p$--$T$ phase diagram of BFCO$_{0.5}$. Phase transition points were determined under the condition $\Delta G (p,T_{t}) 
	\equiv G_{\cal{T}} (p,T_{t}) - G_{\cal{R}} (p,T_{t}) = 0$.} 
\label{fig1}
\end{figure*}

\section*{Results}
{\bf Phase competition in BFCO under pressure.}~At room temperature and zero pressure, BiFe$_{1-x}$Co$_{x}$O$_{3}$ solid 
solutions (BFCO) can be stabilized in two different polymorphs, depending on the relative content of Fe/Co atoms, exhibiting
rhombohedral (${\cal R}$) and tetragonal (${\cal T}$) symmetries \cite{bfco1,bfco2,menendez20}. For relative cobalt contents 
of $0 \le x \lesssim 0.25$, the BFCO ground state is the ${\cal R}$ phase, which is analogous to the ground state of 
bulk BiFeO$_{3}$ \cite{bfco1,bfco2,menendez20,cazorla13}. This rhombohedral phase presents an electric polarization of 
$60$--$80$~$\mu$C~cm$^{-2}$ that is oriented along the pseudocubic direction $[111]$ (Fig.~\ref{fig1}a) and G-type 
antiferromagnetic spin ordering (AFM-G, the net magnetic moment of each transition metal ion is antiparallel to those of its 
six first nearest neighbours). For larger relative cobalt contents, $0.25 < x$, the BFCO ground state corresponds to the 
${\cal T}$ phase, which is analogous to the ground state of bulk BiCoO$_{3}$ \cite{bfco1,bfco2,menendez20,cazorla17,cazorla18}. 
This tetragonal phase presents a giant electric polarization of $160$--$180$~$\mu$C/cm$^{2}$ oriented along the pseudocubic 
direction $[001]$ (Fig.~\ref{fig1}a), hence sometimes it is referred to as ``supertetragonal'', and C-type antiferromagnetic 
spin ordering (AFM-C, the net magnetic moment of each transition metal ion is parallel to those of its two first nearest 
neighbours located along the polar axis and antiparallel to those of its other four first nearest neighbours). 

\begin{figure*}[t]
\centerline
        {\includegraphics[width=0.90\linewidth]{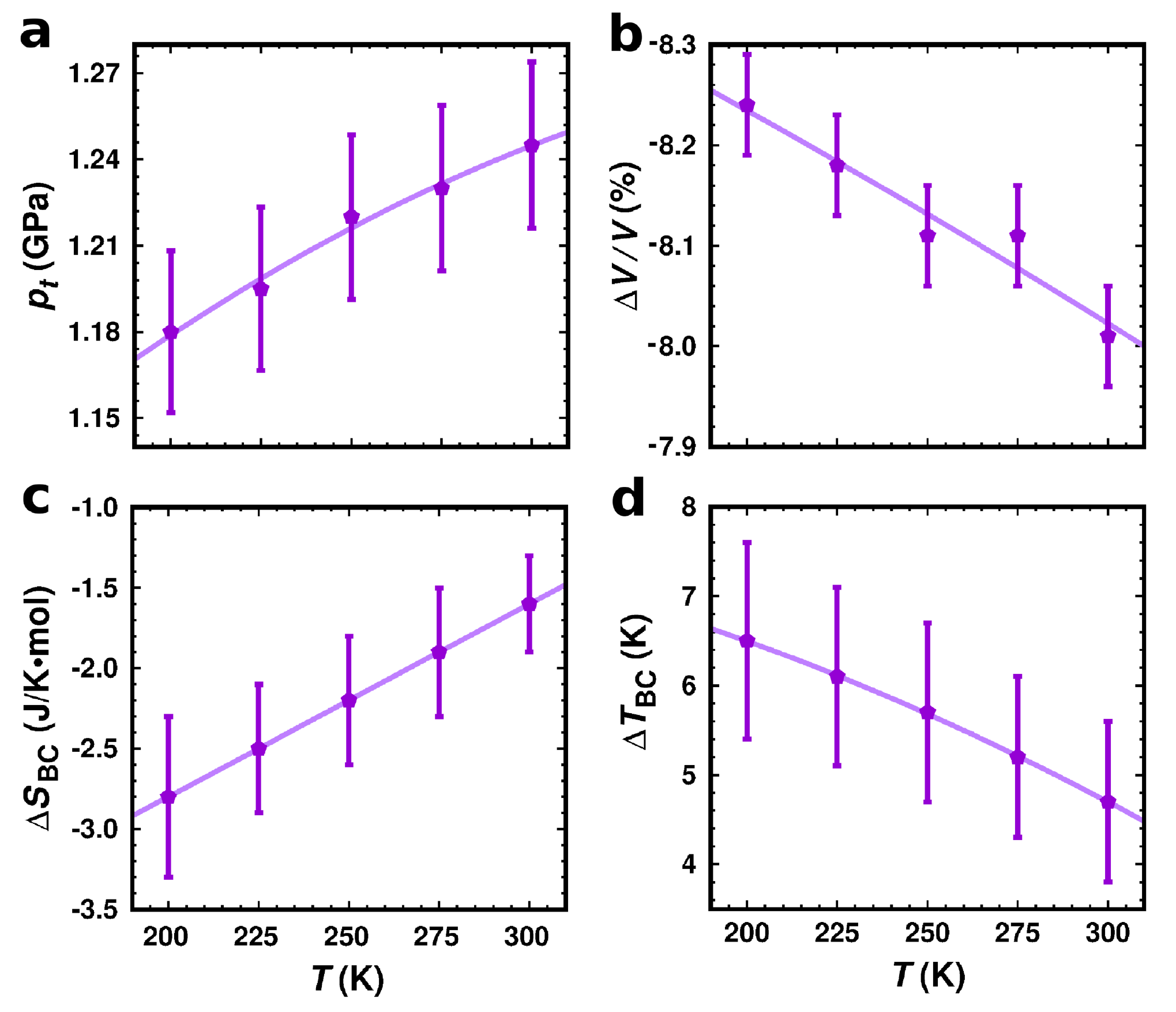}}
        \caption{{\bf Barocaloric descriptors of BFCO$_{0.5}$ estimated with DFT-based first-principles methods.}
        {\bf a}~$\cal{T} \to \cal{R}$ phase transition pressure expressed as a function of temperature.
        {\bf b}~Relative volume change occurring during the $p$-induced $\cal{T} \to \cal{R}$ phase transition as
        referred to that of the tetragonal phase. {\bf c}~Barocaloric isothermal entropy change, $\Delta S_{\rm BC}$,
        expressed as a function of temperature. {\bf d}~Barocaloric adiabatic temperature change, $\Delta T_{\rm BC}$,
        expressed as a function of temperature. Both $\Delta S_{\rm BC}$ and $\Delta T_{\rm BC}$ were estimated
        indirectly by using the Clausius-Clapeyron relation (Methods). Solid lines in the figure are simple eye-guides.}
\label{fig2}
\end{figure*}

Under increasing temperature and for relative cobalt contents of $x \lesssim 0.25$, the supertetragonal ${\cal T}$ phase can
be stabilized over the ${\cal R}$ phase owing to its larger vibrational entropy \cite{bfco1,bfco2,menendez20}. Such a $T$-induced
phase transition clearly is of first-order type (or discontinuous) since the volume change associated to it is huge
($\sim 10$\%). To the best of our knowledge, there are not experimental studies on BFCO under pressure. Here, we amend for 
such a lack of information by carrying out accurate first-principles calculations based on density functional theory (DFT, 
Methods) \cite{menendez20,cazorla17d}. Figure~\ref{fig1}b shows the estimated hydrostatic pressure that is necessary to drive 
the ${\cal T} \to {\cal R}$ phase transition at low temperatures (i.e., disregarding entropy and also likely quantum 
nuclear effects) and for compositions in the interval $0.25 \le x \le 0.50$. This transition pressure is found to 
steadily, and significantly, decrease under increasing Fe content. For instance, $p_{t}$ amounts to $1.4$~GPa at $x = 0.50$ 
and to $0.3$~GPa at $x = 0.25$. As expected, the closer the cobalt content is to the ${\cal T}$--${\cal R}$ morphotropic 
phase boundary ($x_{c} \approx 0.25$), the easier results to switch from the supertetragonal to the rhombohedral phase with 
pressure. 

Simulating temperature effects in materials with first-principles methods is computationally very intensive and laborious. 
However, temperature effects are critical for the assessment of possible caloric phenomena hence cannot be neglected in 
the present study. We employed the quasi-harmonic approximation (QHA) \cite{menendez20,cazorla17d} to calculate \emph{ab initio} 
Gibbs free energies for BFCO in the ${\cal T}$ and ${\cal R}$ phases over broad pressure, temperature and electric field 
conditions, thus allowing for the estimation of barocaloric and electrocaloric effects (Methods). 

Figure~\ref{fig1}c shows the $p$--$T$ phase diagram calculated for BFCO at a composition of $x = 0.50$, hereafter referred 
to as BFCO$_{0.5}$. Therein, it is appreciated that $p_{t}$ consistenly increases under raising temperature, reaching a value 
of $1.24$~GPa at room temperature. In spite of such a relatively large pressure, in what follows we present multicaloric results 
obtained for bulk BFCO$_{0.5}$ at and near room temperature since from a computational point of view this solid solution is highly 
affordable (i.e., the size of the corresponding simulation cells are among the smallest thus making the QHA free energy calculations 
feasible). In practice, much smaller pressures of the order of $0.1$~GPa can be attained by reducing the relative content of Co 
ions (Fig.~\ref{fig1}b) without significantly affecting the main conclusions presented in the next sections. 
\\

\begin{figure*}[t]
\centerline
        {\includegraphics[width=1.00\linewidth]{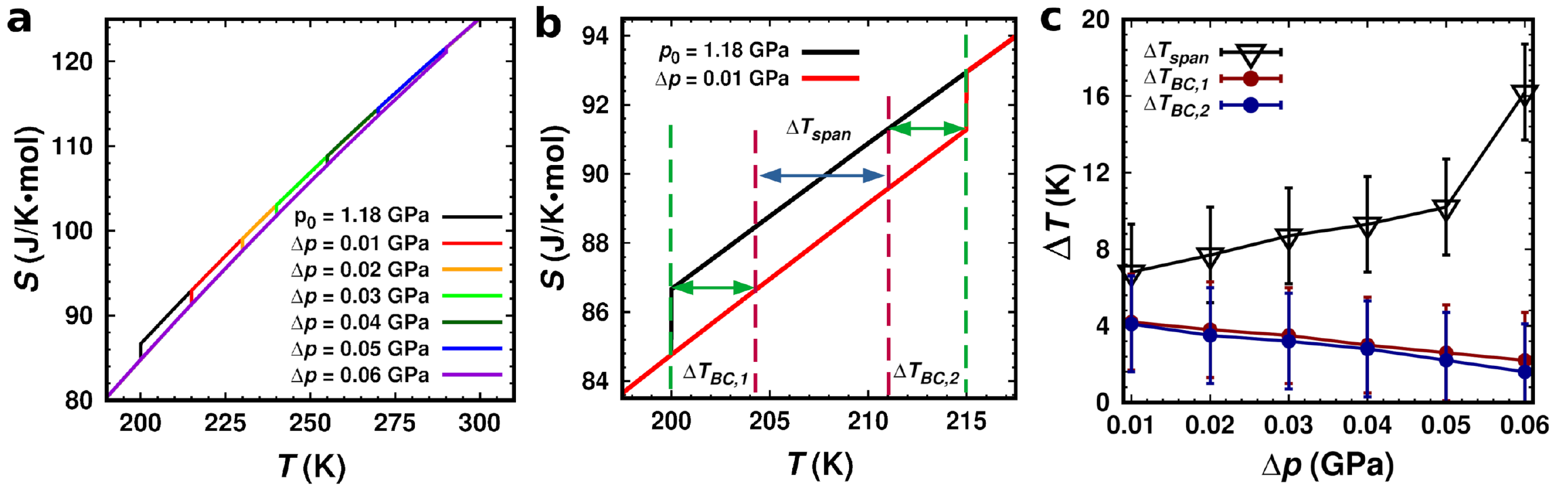}}
        \caption{{\bf Barocaloric performance of BFCO$_{0.5}$ directly estimated with DFT-based first-principles methods.}
	{\bf a}~Entropy curves expressed as a function of temperature and applied pressure shift, $\Delta p \equiv p - p_{0}$.
        {\bf b}~Direct estimation of the adiabatic temperature change, $\Delta T_{\rm BC}$, and temperature span increment,
        $\Delta T_{\rm span}$. The latter quantity is calculated among consecutive pressure shifts of $0.01$~GPa, hence for
        a total pressure shift of $\Delta p = \sum_{i} \Delta p_{i}$ the corresponding temperature span is $T_{\rm span} = 
        \sum_{i} \Delta T_{{\rm span},i}$. {\bf c}~Barocaloric descriptors expressed as a function of the applied pressure
        shift.}
\label{fig3}
\end{figure*}

{\bf Barocaloric performance of BFCO$_{0.5}$.}~We start by analyzing the barocaloric effects induced by hydrostatic pressure 
in bulk BFCO$_{0.5}$ in the absence of electric fields. Figures~\ref{fig2}a-b show the compression required to induce the $\cal{T} 
\to \cal{R}$ phase transition as a function of temperature, $p_{t}$, and the accompanying relative volume change. The estimated 
phase transition volume change is negative and very large as it amounts to $\sim 8$\% in absolute value. Such a huge relative 
volume change augurs a large phase transition entropy change, as it can be inferred from the Clausius-Clapeyron relation 
$\Delta S_{t} = \Delta V \cdot \frac{d{p}_{t}}{dT}$. However, after doing the calculations and assuming that $\Delta S_{\rm BC}
\approx \Delta S_{t}$ (Methods), it was found that the ensuing barocaloric isothermal entropy shifts were actually quite modest 
(Fig.~\ref{fig2}c). For instance, at room temperature we obtained $|\Delta S_{\rm BC}| = 1.7$~J~K$^{-1}$~mol$^{-1}$ 
($5.4$~J~K$^{-1}$~kg$^{-1}$), which is about one order of magnitude smaller than the giant barocaloric entropy changes found in 
superionic and plastic crystals ($\sim 100$~J~K$^{-1}$~kg$^{-1}$) 
\cite{aznar17,cazorla16,cazorla17b,cazorla18b,lloveras19,li19,cazorla19b,sau21,cazorla19}. 
Under decreasing temperature, $|\Delta S_{\rm BC}|$ slightly increases (e.g., $2.8$~J~K$^{-1}$~mol$^{-1}$ at $T = 200$~K) however 
the estimated values still are quite reduced. The reason for these outcomes is that $p_{t}$ barely changes with temperature in
the explored thermodynamic range (i.e., the temperature derivative of the phase transition pressure amounts only to 
$\sim 10^{-3}$~GPa~K$^{-1}$, Fig.~\ref{fig2}a).

The revealed minute $T$-induced $p_{t}$ variation, on the other hand, implies sizeable changes in the phase transition temperature, 
$T_{t}$, induced by small pressure shifts (since $d{T}_{t}/dp = \left[d{p}_{t}/dT\right]^{-1}$), thus suggesting possibly 
large barocaloric thermal shifts in bulk BFCO$_{0.5}$. Figure~\ref{fig2}d shows the barocaloric adiabatic temperature changes, 
$\Delta T_{\rm BC}$, estimated as a function of temperature (Methods). At room temperature ($T = 200$~K), $\Delta T_{\rm BC}$ 
was found to amount to $4.7$~K ($6.5$~K) which, although it cannot rival with the barocaloric adiabatic temperature changes 
reported for superionic and plastic crystals ($\sim 10$~K) \cite{aznar17,cazorla16,cazorla17b,cazorla18b,lloveras19,li19,cazorla19b,sau21,cazorla19}, 
it shows promise in the context of electrocaloric effects ($\sim 1$--$10$~K).

\begin{figure*}[t]
\centerline
        {\includegraphics[width=1.00\linewidth]{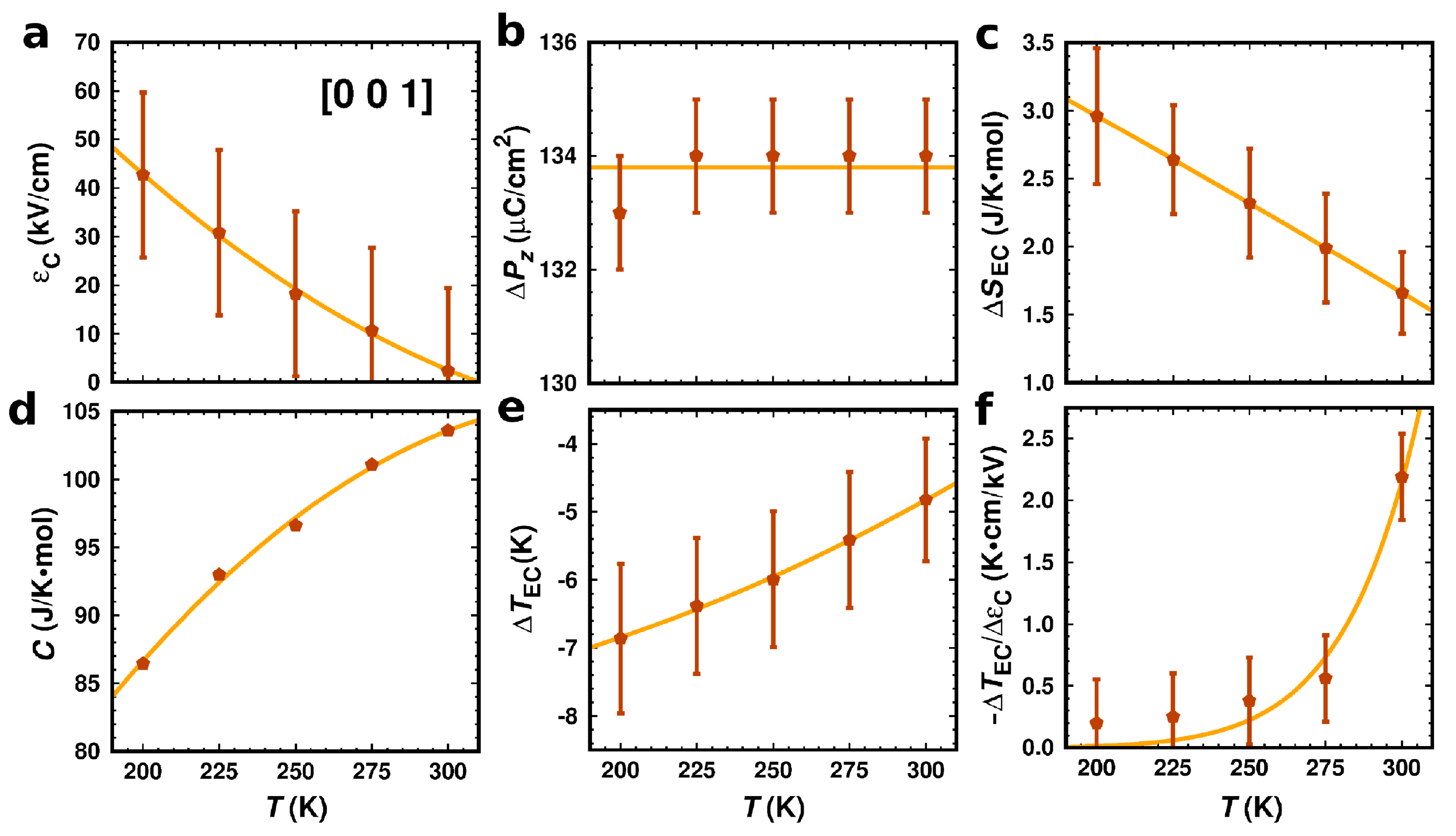}}
        \caption{{\bf Electrocaloric performance of BFCO$_{0.5}$ estimated with DFT-based first-principles methods
        at a fixed pressure of $1.25$~GPa.}
        {\bf a}~Critical electric field applied along the [001] direction inducing the $\cal{R} \to \cal{T}$
        phase transition.
	{\bf b}~Electric polarization change along the [001] direction occurring during the ${\cal E}$-induced $\cal{R} 
	\to \cal{T}$ phase transition.
        {\bf c}~Electrocaloric isothermal entropy change, $\Delta S_{\rm EC}$, calculated for the ${\cal E}$-induced 
	$\cal{R} \to \cal{T}$ phase transformation in compressed BFCO$_{0.5}$.
        {\bf d}~Heat capacity of compressed BFCO$_{0.5}$.
        {\bf e}~Electrocaloric adiabatic temperature change, $\Delta T_{\rm EC}$, calculated for the ${\cal E}$-induced 
        $\cal{R} \to \cal{T}$ phase transformation in compressed BFCO$_{0.5}$. 
        {\bf f}~Electrocaloric strength of compressed BFCO$_{0.5}$. Both $\Delta S_{\rm EC}$ and $\Delta T_{\rm EC}$ were 
	estimated indirectly by using the Clausius-Clapeyron relation (Methods). Solid lines in the figure are simple 
        eye-guides.}
\label{fig4}
\end{figure*}

The barocaloric results presented above were obtained with the indirect Clausius-Clayperon (CC) method, which is not exact \cite{aznar17}. 
Aimed to assess the extent of the employed approximations, we mimicked with theory quasi-direct barocaloric experiments \cite{aznar17,lloveras19}
in which entropy curves are estimated as a function of pressure and temperature and from which $\Delta S_{\rm BC}$ and $\Delta T_{\rm BC}$ 
can be straightforwardly determined (Figs.~\ref{fig3}a-b) \cite{cazorla19}. Moreover, with this quasi-direct estimation approach is 
also possible to determine for a given pressure shift, $\Delta p$, the temperature span, $T_{\rm span}$, over which barocaloric effects 
can be operated (Fig.~\ref{fig3}b). In view of the huge $d{T}_{t}/dp$ of $\sim 10^{3}$~K~GPa$^{-1}$ estimated for BFCO$_{0.5}$, giant 
$T_{\rm span}$ values are anticipated \cite{li20}.

Figure~\ref{fig3}c shows the results of our quasi-direct barocaloric descriptor estimations. At room temperature and $T = 200$~K, 
we obtained adiabatic temperature changes of $2.0 \pm 2.5$ and $4.0 \pm 2.5$~K, respectively. Within the numerical uncertainties, 
these results are compatible with our previous estimations obtained with the CC method; however, it goes without saying that the 
reported error bars are unacceptably too large. The reasons for the relatively huge numerical uncertainties on $\Delta T_{\rm BC}$ 
are the small $\Delta S_{t}$ and great $p$-induced $T_{t}$ shifts involved in its quasi-direct estimation (Fig.~\ref{fig3}a). Thus, 
unfortunately, in the present case it is not possible to discern the actual precision of the barocaloric adiabatic temperature 
changes obtained with the approximate CC method. Nevertheless, the estimation of $T_{\rm span}$ is still possible given its noticeably 
large size (Figs.~\ref{fig3}a-b). By considering an outset compression of $1.18$~GPa, we obtained $T_{\rm span} \approx 60$~K for a 
small pressure shift of $0.06$~GPa (calculated by adding up all the $\Delta T_{\rm span}$ increments shown in Fig.~\ref{fig3}c). 
This result is very encouraging since it indicates that, in spite of the relative smallness of $\Delta S_{\rm BC}$ and $\Delta T_{\rm BC}$, 
barocaloric effects in BFCO$_{0.5}$ could be operated over unusually ample temperature ranges. 
\\

{\bf Electrocaloric performance of pressurized BFCO$_{0.5}$.}~The electric polarization, $P$, of BFCO$_{0.5}$ in the ${\cal R}$ 
and ${\cal T}$ phases are significantly different; for instance, $P$ in the supertetragonal phase is more than two times larger 
than that in the rhombohedral phase \cite{menendez20}, adding up to polarization module differences of $> 100$~$\mu$C~cm$^{-2}$ 
(Fig.~\ref{fig4}b). Such a huge electric polarization disparity seems very promising from an electrocaloric (EC) point of view, 
as it can be inferred from the electric Clausius-Clapeyron relation $\Delta S_{t} = \Delta \bm{P} \cdot \frac{d\bm{{\cal E}_{c}}}{dT}$, 
where $\Delta S_{t}$ represents the entropy change associated to the field-induced phase transition and ${\cal E}_{c}$ 
the necessary electric field to switch from the ${\cal R}$ to the ${\cal T}$ phase. Figure~\ref{fig4}a shows the ${\cal E}_{c}$ 
estimated for a fixed pressure of $1.25$~GPa as a function of temperature (Methods), which has been selected to ensure proper 
stabilization of the ${\cal R}$ phase under conditions $T \le 300$~K. As clearly appreciated therein, the critical electric field 
steadily decreases under increasing temperature, ranging from $43$~kV~cm$^{-1}$ at $200$~K to $\approx 2$~kV~cm$^{-1}$ at room 
temperature.

Figures~\ref{fig4}c-e show the electrocaloric isothermal entropy and adiabatic temperature changes, $\Delta S_{\rm EC}$ 
and $\Delta T_{\rm EC}$, estimated for compressed BFCO$_{0.5}$ using the indirect CC approach (Methods). In this case, 
the sign of the EC descriptors indicate that the caloric effect is inverse (i.e., $\Delta T < 0$), which follows from the 
fact that the high-entropy phase ${\cal T}$ presenting the largest electric polarization is stabilized via application of 
the external electric bias. As expected, the size and temperature dependence of $|\Delta S_{\rm EC}|$ and $|\Delta T_{\rm EC}|$,
which are directly related through the temperature and heat capacity (Fig.~\ref{fig4}d, Methods), are very much similar 
to those of $|\Delta S_{\rm BC}|$ and $|\Delta T_{\rm BC}|$ since the underlying phase transitions are equivalent.
For instance, at $T = 200$~K we estimated an electrocaloric adiabatic temperature change of $-6.9$~K and at room 
temperature of $-4.8$~K, to be compared with the analogous barocaloric shifts of $+6.5$ and $+4.7$~K. These $\Delta T_{\rm EC}$ 
values are very much promising, specially when considering the small size of the required driving electric fields 
(that is, ${\cal E}_{c} \sim 1$--$10$~kV~cm$^{-1}$).

Figure~\ref{fig4}f encloses results for the electrocaloric strength of BFCO$_{0.5}$, $\Lambda_{\rm EC}$, expressed as a 
function of temperature; this quantity is defined like the ratio of $\Delta T_{\rm EC}$ by the corresponding electric bias. 
At $T = 200$~K, the attained adiabatic temperature change is highest however the required switching electric field is 
also largest, thus the resulting electrocaloric strength is smaller than obtained at higher temperatures. Still, the 
calculated $\Lambda_{\rm EC}$ amounting to $0.2$~K~cm~kV$^{-1}$ is already comparable to the record experimental values 
reported for oxide and hybrid organic-inorganic perovskites \cite{bto1,bto2,ec7}. Remarkably, under increasing temperature 
the electrocaloric strength of BFCO$_{0.5}$ noticeably increases reaching a maximum, and colossal, value of $2.2$~K~cm~kV$^{-1}$ 
at $T = 300$~K. These figures will be put into context in a next section; in what follows, we explain how the dual
response of BFCO$_{0.5}$ to mechanical and electric stimuli may be exploited in practical solid-state cooling cycles. 
\\

\begin{figure*}[t]
\centerline
        {\includegraphics[width=1.00\linewidth]{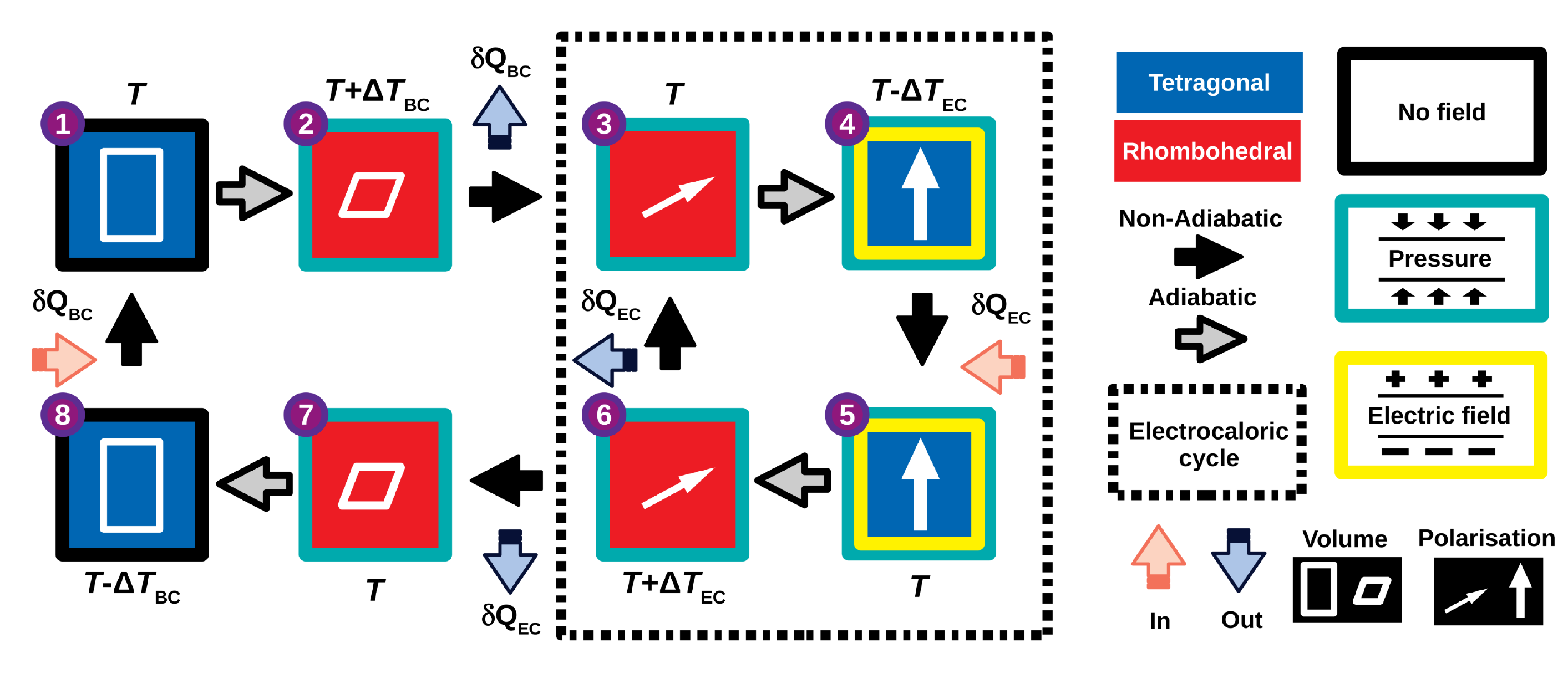}}
	\caption{{\bf Sketch of the proposed $p$--${\cal E}$ multicaloric cycle for enhancement of the electrocaloric strength.} 
	(1)~The multiferroic compound BFCO is at equilibrium in the $\cal{T}$ phase at temperature $T$. 
	(2)~Hydrostatic pressure is adiabatically applied on BFCO so that it transforms into the $\cal{R}$ 
	phase and experiences a temperature increase of $\Delta T_{\rm BC}$.
	(3)~Heat, $\delta Q_{\rm BC}$, is released to the ambient and the initial temperature is restored; 
	compressed BFCO remains in the $\cal{R}$ phase.
	(4)~An electric field is adiabatically applied on compressed BFCO so that it transforms into the $\cal{T}$ 
	phase and experiences a temperature decrease of $|\Delta T_{\rm EC}|$.
	(5)~Heat, $\delta Q_{\rm EC}$, is absorbed by the system and the initial temperature is restored; compressed 
	and electrically biased BFCO remains in the $\cal{T}$ phase.
	(6)~The electric field is adiabatically removed from compressed BFCO, thus it transforms into the 
	$\cal{R}$ phase and experiences a temperature increase of $|\Delta T_{\rm EC}|$. 
	(7)~Heat, $\delta Q_{\rm EC}$, is released to the ambient and the initial temperature is restored; compressed 
	BFCO remains in the $\cal{R}$ phase. The state reached in this step is equivalent to that in step (3), 
	thus one can repeatedly run the electrocaloric subcycle (3)-(4)-(5)-(6) entailing application and removal 
	of an electric bias under fixed hydrostatic pressure (dashed lines).
	(8)~Hydrostatic pressure is adiabatically released from BFCO so that it transforms into the $\cal{T}$
        phase and experiences a temperature decrease of $\Delta T_{\rm BC}$. 
	Heat, $\delta Q_{\rm BC}$, is absorbed by the system and the starting temperature is restored, realizing an
	entire multicaloric (1)-(8) cycle.}
\label{fig5}
\end{figure*}

\begin{table*}
\centering
\begin{tabular}{c c c c c c}
\hline
\hline
$ $ & $ $ & $ $ & $ $ & $ $ & $ $ \\
$ $ & \quad $T$ \quad & \quad  $\varepsilon_{c}$ \quad & \quad $\Delta T_{\rm EC}$ \quad & \quad $|\Delta T_{\rm EC}|/\varepsilon_{c}$ \quad & \quad ${\rm References}$ \qquad \\
        $ $ \quad & \quad ${\rm (K)}$ \quad & \quad ${\rm (kV~cm^{-1})}$ \quad & \quad ${\rm (K)}$ \quad & \quad ${\rm (K~cm~kV^{-1})}$ \quad & $ $ \\
$ $ & $ $ & $ $ & $ $ & $ $ & $ $ \\
\hline
$ $ & $ $ & $ $ & $ $ & $ $ & $ $  \\
 Y-HfO$_{2}$                                   &   358 &        3500 & 24.8 &   0.01  & [\onlinecite{ec0}] \\
 0.93PMN-0.07PT                                &   298 &         723 &  9.0 &   0.01  & [\onlinecite{ec2}] \\
 (NH$_{4}$)$_{2}$SO$_{4}$                      &   220 &         400 &  4.5 &   0.01  & [\onlinecite{as2}] \\
 Terpolymer/PMN-PT                             &   303 &        1800 & 31.0 &   0.02  & [\onlinecite{ec3}] \\
 Ba$_{0.65}$Sr$_{0.35}$TiO$_{3}$               &   293 &         130 &  3.1 &   0.02  & [\onlinecite{ec5}] \\
 BaZr$_{0.2}$Ti$_{0.8}$O$_{3}$                 &   313 &         145 &  4.5 &   0.03  & [\onlinecite{ec4}] \\
 BNBT-BCZT                                     &   370 &         620 & 23.0 &   0.04  & [\onlinecite{ec6}] \\
 PbZr$_{0.46}$Sn$_{0.46}$Ti$_{0.08}$O$_{3}$    &   317 &          30 &  1.6 &   0.05  & [\onlinecite{ec1}] \\
 BaTiO$_{3}$                                   &   400 &         4.0 &  0.9 &   0.23  & [\onlinecite{bto1}]~[\onlinecite{bto2}] \\
 ((CH$_{3}$)$_{2}$CHCH$_{2}$NH$_{3}$)$_{2}$PbCl$_{4}$  &         302 &  30  &   11.1  &  0.37  & [\onlinecite{ec7}] \\
 BFCO$_{0.5}$~(pressurized)                    &   300 &         2.2 & -4.8 &   2.18  & This work \\
$ $ & $ $ & $ $ & $ $ & $ $ & $ $ \\
\hline
\hline
\end{tabular}
\label{tab:mcperform}
        \caption{{\bf Electrocaloric performance of several ferroelectric materials at or near room temperature.}
          The electrocaloric strength of compressed BFCO$_{0.5}$ is significantly larger than those of
          other uncompressed ferroelectric compounds.
         }
\end{table*}

{\bf Proposed $p$--$E$ multicaloric cycle.}~Single stimulus solid-state cooling cycles typically consists of four 
thermodynamic steps, two involving the adiabatic switching on and off of the applied external field and the 
other two constant-field heat transfer processes with the environment and the system to be refrigerated 
\cite{cazorla19}. In the present work, we propose an original multi-stimuli solid-state cooling cycle consisting 
of eight thermodynamic steps that has been designed to minimize the applied electric field, thus maximizing
$\Lambda_{\rm EC}$, and with a cumulative multicaloric performance of $|\Delta T_{\rm MC}| = |\Delta T_{\rm BC}| 
+ |\Delta T_{\rm EC}|$ and $|\Delta S_{\rm MC}| = |\Delta S_{\rm BC}| + |\Delta S_{\rm EC}|$. 

Figure~\ref{fig5} sketches the envisaged multi-stimuli solid-state cooling cycle comprising hydrostatic pressure and
electric fields being applied on multiferroic lead-free BFCO solid solutions near room temperature. The cycle starts 
with multiferroic BFCO in the supertetragonal $\cal{T}$ phase at temperature $T$. Subsequently, hydrostatic pressure 
is adiabatically applied on BFCO so that it transforms into the $\cal{R}$ phase and experiences a temperature increase 
of $|\Delta T_{\rm BC}|$. In the third step, heat is released to the ambient, $\delta Q_{\rm BC}$, and the initial 
temperature of the cycle is restored; compressed BFCO still remains in the $\cal{R}$ phase. Next, an electric field 
is adiabatically applied on compressed BFCO so that it transforms into the $\cal{T}$ phase, thus experiencing a 
temperature decrease of $|\Delta T_{\rm EC}|$. In the fifth step, heat is absorbed by the system, $\delta Q_{\rm EC}$, 
and the initial temperature of the cycle is restored; compressed and electrically biased BFCO remains in the $\cal{T}$ 
phase. Subsequently, the electric field is adiabatically removed thus BFCO transforms into the $\cal{R}$ phase and 
experiences a temperature increase of $|\Delta T_{\rm EC}|$. In the seventh step, heat is released to the ambient,
$\delta Q_{\rm EC}$, and the initial temperature of the cycle is restored; compressed BFCO remains in the $\cal{R}$ 
phase. Finally, hydrostatic pressure is adiabatically released so that BFCO transforms back into the $\cal{T}$ phase 
and experiences a temperature decrease of $|\Delta T_{\rm BC}|$. Heat then is absorbed by the system, $\delta Q_{\rm BC}$, 
and the initial temperature of the cycle is restored, thus completing an entire multi-stimuli cycle.

Upon completion of a multi-stimuli cycle, multiferroic BFCO is able to remove an amount of heat equal to $|\delta Q_{\rm BC}| 
+ |\delta Q_{\rm EC}|$, or equivalently, $T \cdot \left(|\Delta S_{\rm BC}| + |\Delta S_{\rm EC}|\right)$, from the targeted 
system to be refrigerated and release it to the ambient (thus cooling it down). The described multi-stimuli cycle 
lends itself to several useful variations. For instance, the state reached in the seventh step is thermodynamically equivalent 
to that attained in the third; therefore, one could recursively perform the electrocaloric subcycle consisting of steps 
(3)--(4)--(5)--(6) which entails application and removal of an electric bias under fixed hydrostatic pressure (dashed 
lines in Fig.~\ref{fig5}). Likewise, if the multi-stimuli cooling cycle starts with multiferroic BFCO in the rhombohedral 
$\cal{R}$ phase instead of the $\cal{T}$ phase, due to some composition synthesis constraints, for example, then the 
sequential application of hydrostatic pressure and electric field explained above needs to be swapped.

\section*{Discussion}
Table~I summarizes some representative materials for which EC effects occurring at or near room temperature have been 
experimentally measured and reported in the literature. The selected compounds belong to three different families of 
ferroelectric materials, namely, oxides (e.g., HfO$_{2}$ and BaTiO$_{3}$), hybrid organic-inorganic perovskites 
([(CH$_{3}$)$_{2}$CHCH$_{2}$NH$_{3}$]$_{2}$PbCl$_{4}$) and polymers (Terpolymer). In terms of largest $|\Delta T_{\rm EC}|$,
the oxides Y-HfO$_{2}$ \cite{ec0} and BNBT-BCZT \cite{ec6} and the elastomer Terpolymer/PMN-PT \cite{ec3} emerge as 
the most promising since they display colossal values of $20$--$30$~K. Nevertheless, these record materials require of 
quite large electric fields to realize their full EC potential (${\cal E}_{c} \sim 10^{3}$~kV~cm$^{-1}$), hence with 
no exception their associated electrocaloric strengths turn out to be quite mediocre, namely, $\Lambda_{\rm EC} \sim 
0.01$~K~cm~kV$^{-1}$.

Ferroelectric materials exhibiting moderate or even small $|\Delta T_{\rm EC}|$ but attained under smaller electric 
fields (${\cal E}_{c} \sim 10$~kV~cm$^{-1}$), on the other hand, become the clear winners in terms of largest 
$\Lambda_{\rm EC}$. For instance, the archetypal perovskite oxide BaTiO$_{3}$ renders an adiabatic temperature change
of roughly $1$~K driven by a minute electric field of $4$~kV~cm$^{-1}$, thus leading to a huge electrocaloric strength 
of $0.23$~K~cm~kV$^{-1}$ \cite{bto1,bto2}. Likewise, the hybrid organic-inorganic perovskite 
[(CH$_{3}$)$_{2}$CHCH$_{2}$NH$_{3}$]$_{2}$PbCl$_{4}$ holds the record $\Lambda_{\rm EC}$ value of $0.37$~K~cm~kV$^{-1}$, 
which results from a small electric field of $30$~kV~cm$^{-1}$ and an adiabatic temperature change of $11.1$~K \cite{ec7}. 
It is worth noting that all these figures correspond to experimental data.  

Table~I encloses also the EC results that we have theoretically estimated in this study for pressurized BFCO$_{0.5}$
at room temperature. According to our QHA-DFT calculations, compressed multiferroic BFCO solid solutions have the potential 
to surpass all previously known EC materials in terms of largest $\Lambda_{\rm EC}$. In particular, we predict an outstanding
electrocaloric strength of $2.18$~K~cm~kV$^{-1}$ that arises from an adiabatic temperature change of $4.8$~K and an electric 
bias of $\approx 2$~kV~cm$^{-1}$. This theoretically estimated $\Lambda_{\rm EC}$ value is from one to two orders of 
magnitude larger than those experimentally measured in uncompressed ferroelectrics. The key mechanism in achieving such 
a colossal figure is to employ an ancillary field, in our case hydrostatic pressure, to bring the system towards the verge 
of a ferroelectric phase transition so that it is possible to drive it with a minuscule electric field.

In the specific case considered here, the pressure required to achieve the colossal $\Lambda_{\rm EC}$ value of 
$2.18$~K~cm~kV$^{-1}$ is higher than $1$~GPa. Obviously, this compression is too large to be considered for practical 
applications. Nevertheless, as it was argued at the begining of the Results section, it is possible to significantly 
reduce the size of this ancillary pressure to the order of $0.1$~GPa by decreasing the relative content of 
cobalt ions down to the critical composition of $\approx 0.25$. Moreover, the $\Lambda_{\rm EC}$ enhancement approach 
proposed in this study, and theoretically demonstrated for BFCO$_{0.5}$, in principle should be generalizable to 
many other well-known EC materials since most of them are responsive to pressure as well (even though the magnitude of 
the resulting BC effects may be quite small in comparison to those achieved in state-of-the-art barocaloric materials). Take 
the archetypal ferroelectric compound BaTiO$_{3}$ as an example. The ferro- to paraelectric phase transition temperature 
in this material can be effectively shifted with pressure, namely, $d{T}_{t}/dp \approx -25$~K~GPa$^{-1}$ \cite{hayward02}, 
thus its room-temperature EC performance could be potentially improved with our proposed strategy. Finally, to mention that
recent developments in the synthesis of ferroelectric membranes and thin films may also allow for the enhancement of
the $\Lambda_{\rm EC}$ figure-of-merit by combining electric fields with other types of mechanical stimuli like uniaxial 
\cite{zang22} and biaxial \cite{liu16b} stress. 
\\

In conclusion, we have proposed a new strategy for the enhancement of the electrocaloric strength of ferroelectric 
materials that consists in concertedly applying pressure and electric fields. We have theoretically proved our new
concept on multifunctional BFCO solid solutions, an intriguing family of compounds displaying a discontinuous phase 
transition between two multiferroic states. In particular, for compressed BFCO$_{0.5}$ we estimated a record 
$\Lambda_{\rm EC}$ parameter of $2.18$~K~cm~kV$^{-1}$ at room temperature resulting from an adiabatic temperature 
change of $4.8$~K and an electric bias of $\approx 2$~kV~cm$^{-1}$. This electrocaloric strength turns out to be 
colossal since it is from one to two orders of magnitude larger than those experimentally measured in uncompressed 
ferroelectrics. The demonstrated $\Lambda_{\rm EC}$ enhancement strategy can be applied to other types of ferroelectric 
materials, not necessarily magnetic, and be modified at convenience on the mechanical component. Thus, the combination 
of multiple stimuli opens new horizons in the field of caloric materials and solid-state refrigeration by expanding the 
design of possible cooling cycles and boosting current caloric performances. We hope that the present theoretical study 
will motivate new experimental works on the engineering of original and environmentally friendly solid-state cooling 
devices.

\section*{Methods}
Spin-polarized DFT calculations were performed with the generalized gradient approximation proposed by Perdew,
Burke and Ernzerhof (PBE) as it is implemented in the VASP package \cite{vasp,pbe96}. The ``Hubbard-$U$'' scheme 
due to Dudarev \textit{et al.} was employed in the PBE calculations for treating better the Co~(Fe) $3d$ electrons, 
adopting a $U$ value of $6$~($4$)~eV \cite{menendez20,menendez20b,cazorla17,cazorla18,cazorla13}. The ``projected 
augmented wave'' method \cite{bloch94} was used to represent the ionic cores considering the following electronic 
states as valence: Co $4s^{1}3d^{8}$, Fe $3p^{6}4s^{1}3d^{7}$, Bi $6s^{2}5d^{10}6p^{3}$, and O $2s^{2}2p^{4}$. An 
energy cut-off of $800$~eV and a $\Gamma$-centered ${\bf k}$-point grid of $4 \times 6 \times 6$ were employed for 
a $2 \times \sqrt{2} \times \sqrt{2}$ simulation cell containing $20$ atoms \cite{cazorla15}, thus obtaining 
zero-temperature energies converged to within $0.5$~meV/f.u. Geometry relaxations were performed for an atomic force 
threshold of $0.005$~eV$\cdot$\AA$^{-1}$. Electric polarizations were accurately estimated with the hybrid HSE06 
functional \cite{hse06} and the Berry phase formalism \cite{king93,resta94,bellaiche99}. 

\textit{Ab initio} free energies were estimated within the quasi-harmonic approximation (QHA) \cite{cazorla13,cazorla17d,phonopy} 
as a function of $p$ and $T$. Phonon frequencies were calculated with the small displacement method \cite{phonopy}. 
The following technical parameters provided QHA free energies converged to within $5$~meV per formula unit: $160$-atom 
supercells, atomic displacements of $0.01$~\AA, and ${\rm q}$-point grids of $16 \times 16 \times 16$ for integration 
within the first Brillouin zone. The effects of chemical disorder were addressed by generating all possible atomic Co--Fe 
and magnetic spin arrangements (ferromagnetic --FM-- and antiferromagnetic --AFM-- of type A, C, and G) for a $2 \times 
2\sqrt{2} \times \sqrt{2}$ supercell containing $40$ atoms. Quasi-harmonic free energies were calculated only for the 
lowest-energy configurations. Our spin-polarized DFT calculations were performed for bulk BiFe$_{0.5}$Co$_{0.5}$O$_{3}$. 

Within the QHA \cite{cazorla13,cazorla17d,phonopy}, the Gibbs free energy of a given crystal phase, $G_{\rm harm}$, is 
expressed as:
\begin{equation}
G_{\rm harm}(p,T) = E(p) + pV(p,T) + F_{\rm harm}(p,T)~,
\label{eqgibbs}
\end{equation}
where $E$ is the static energy of the system (i.e., as directly obtained from zero-temperature DFT calculations),
$p$ the pressure, $V$ the volume, and $F_{\rm harm}$ the lattice Helmholtz free energy. (The dependence of the
different energy terms on $p$ and $T$ have been explicitly noted.) For given $V$ and $T$, $F_{\rm harm}$ can be
determined with the formula:
\begin{eqnarray}
        F_{\rm harm}(V,T) & = & \frac{1}{N_{q}}~k_{B} T \times \nonumber \\
			  &   & \sum_{{\bf q}s}\ln\left[ 2\sinh \left( \frac{\hbar \omega_{{\bf q}s}}{2k_{B}T} 
			  \right) \right]~,
\label{eqfharm}
\end{eqnarray}
where $\omega_{{\bf q}s}(V)$ are the phonon frequencies obtained at the reciprocal lattice vector ${\bf q}$ and
phonon branch $s$, $N_{q}$ the total number of wave vectors used for integration in the Brillouin zone, and 
$k_{B}$ the Boltzmann constant. Meanwhile, the hydrostatic pressure $p$ is calculated via the expression:
\begin{equation}
p(V,T) = -\frac{\partial \left[E(V) + F_{\rm harm}(V,T)\right]}{\partial V}~, 
\label{eqpress}
\end{equation}
which numerically allows to determine $V(p,T)$. Thus, by performing $E$ and $\omega_{{\bf q}s}$ DFT calculations
for a set of $V$ points, over which interpolation is applied to describe $F_{\rm harm}$ and $p$ continuously, and
by using Eqs.~(\ref{eqgibbs})--(\ref{eqpress}), it is possible to estimate $G_{\rm harm}(p,T)$. To quantify the 
temperature at which the $\cal{T} \leftrightarrow \cal{R}$ phase transition occurs at a given pressure, 
$T_{t}$, the condition $\Delta G_{\rm harm}(p, T_{t}) \equiv G_{\rm harm}^{\cal{T}}(p, T_{t}) 
- G_{\rm harm}^{\cal{R}}(p, T_{t}) = 0$ was employed. 

Likewise, the entropy of the crystal can be obtained through the expression:
\begin{equation}
	S(V,T) = -\left( \frac{\partial F_{\rm harm}}{\partial T} \right)_{V}~,
\label{entropy}
\end{equation}
and the heat capacity like:
\begin{eqnarray}
	C(V,T) & = & k_{B} \sum_{{\bf q}s} \left( \frac{\hbar \omega_{{\bf q}s}}{k_{B}T} \right)^{2} \times \nonumber \\ 
	       &   & \frac{\exp{\left( \hbar \omega_{{\bf q}s} / k_{B}T \right)}}{\left[\exp{\left( 
		 \hbar \omega_{{\bf q}s} / k_{B}T \right)}  -1 \right]^{2}}~. 
\label{heatcap}
\end{eqnarray}
Through the knowledge of $V(p,T)$ and Eqs.~(\ref{eqfharm})--(\ref{heatcap}), then it is possible to determine $S(p,T)$ 
and $C(p,T)$.

In the absence of electric fields, the isothermal entropy change associated to barocaloric effects was approximately 
estimated with the Clausius-Clapeyron (CC) method like \cite{sau21}:
\begin{equation}
\Delta S_{\rm BC}(p, T) = \Delta V \cdot \frac{d{p}_{t}}{dT}~,
\label{eq4}
\end{equation}
where $\Delta V$ is the change in volume occurring during the phase transition and $p_{t} (T)$ the critical pressure. 
Likewise, the corresponding adiabatic temperature change can be approximated with the expression \cite{manosa17}:
\begin{equation}
\Delta T_{\rm BC}(p, T) = -\frac{T}{C} \cdot \Delta S_{\rm BC}(p, T)~, 
\label{eq5}
\end{equation}
where $C(T)$ is the heat capacity of the system at zero pressure.

In the presence of electric fields, and assuming zero pressure, the thermodynamic potential that describes 
a particular phase is the Gibbs free energy defined as $G_{\rm harm} = E - \bm{{\cal E}} \cdot \bm{P} +
 F_{\rm harm}$, where $E$ and $F_{\rm harm}$ are the same terms that appear in Eq.~(\ref{eqgibbs}), $\bm{P}$ the electric
 polarization and $\bm{{\cal E}}$ the applied electric field. In this case, the thermodynamic condition that determines a 
 ${\cal E}$--induced phase transition is $G_{\rm harm}^{\cal T} (T, {\cal E}_{c}) = G_{\rm harm}^{\cal R} (T, {\cal E}_{c})$. 
 The value of the corresponding critical electric field then can be estimated like:
\begin{equation}
	{\cal E}_{c} (T) = \frac{\Delta \left(E + F_{\rm harm} (T)\right)}{\Delta P (T)}~,
\label{eq9}
\end{equation}
where $\Delta \left(E + F_{\rm harm}\right)$ is the Helmholtz free energy difference between the two phases, and $\Delta P$ 
the resulting change in the electric polarization along the electric field direction. For $p \neq 0$ conditions, an additional
$p \Delta V$ term should appear in the right-hand side of Eq.~(\ref{eq9}).

Once the value of ${\cal E}_{c}$ and its dependence on temperature are determined through Eq.~(\ref{eq9}), the isothermal 
entropy change associated to electrocaloric effects can be approximately estimated with the CC method like \cite{cazorla18}:
\begin{equation}
\Delta S_{\rm EC}({\cal E}, T) = -\Delta P \cdot \frac{d{\cal E}_{c}}{dT}~.
\label{eq7}
\end{equation}
Likewise, the corresponding adiabatic temperature change was approximated with the expression \cite{manosa17}:
\begin{equation}
\Delta T_{\rm EC}({\cal E}, T) = -\frac{T}{C} \cdot \Delta S_{\rm EC}({\cal E}, T)~, 
\label{eq6}
\end{equation}
where $C(T)$ is the heat capacity of the system at zero electric field.

\section*{ACKNOWLEDGEMENTS}
The authors thank Riccardo Rurali for insightful discussions on the calculation of phonon properties of BFCO solid 
solutions. C.C. acknowledges financial support from the Spanish Ministry of Science, Innovation and Universities under 
the ``Ram\'on y Cajal'' fellowship RYC2018-024947-I and TED2021-130265B-C22 project. Computational support was provided 
by the Red Espa\~nola de Supercomputaci\'on (RES) under the grants FI-2022-1-0006, FI-2022-2-0003 and FI-2022-3-0014.

\end{document}